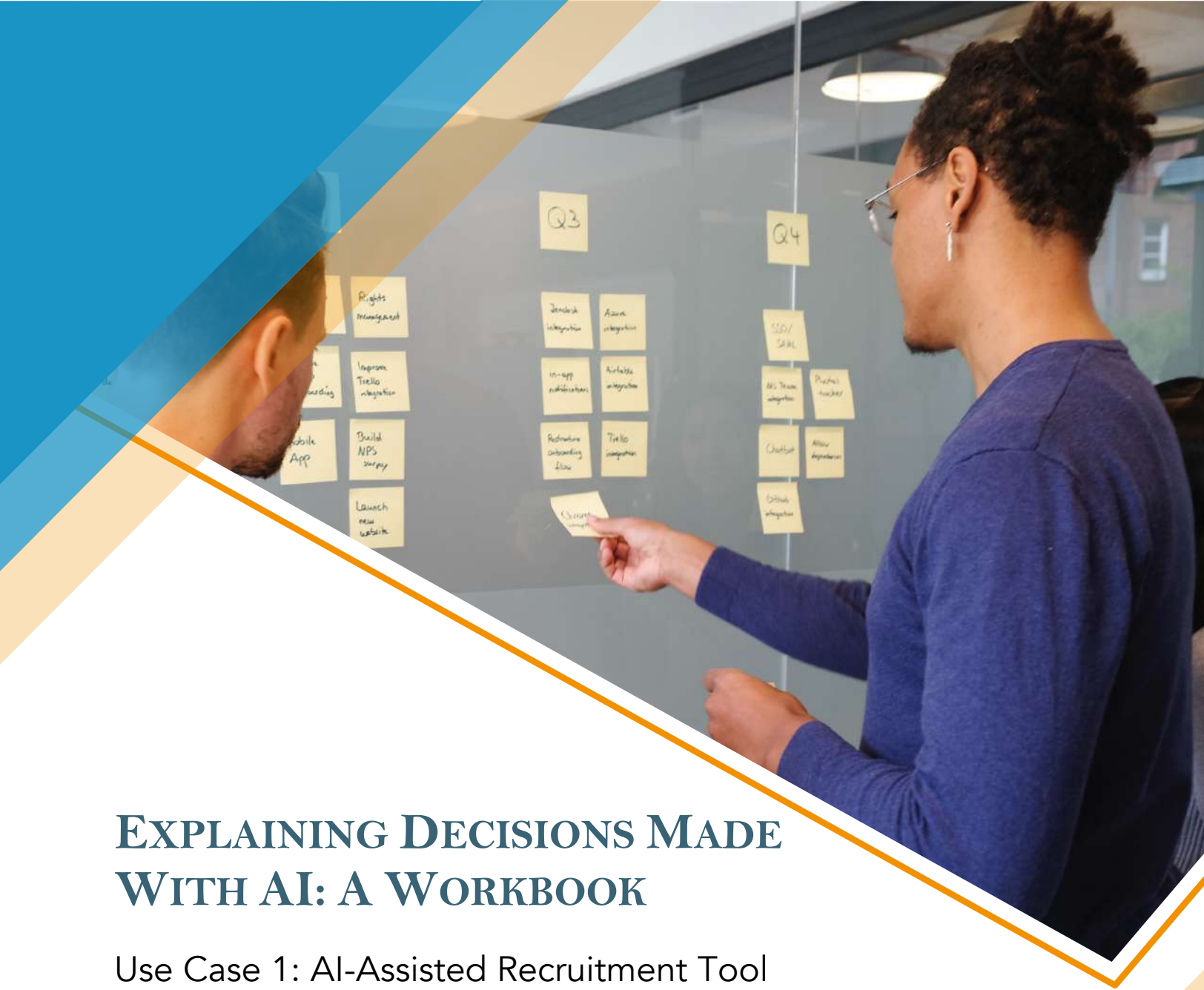

# Explaining Decisions Made With AI: A Workbook

Use Case 1: AI-Assisted Recruitment Tool

Dr David Leslie and Morgan Briggs

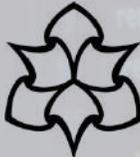

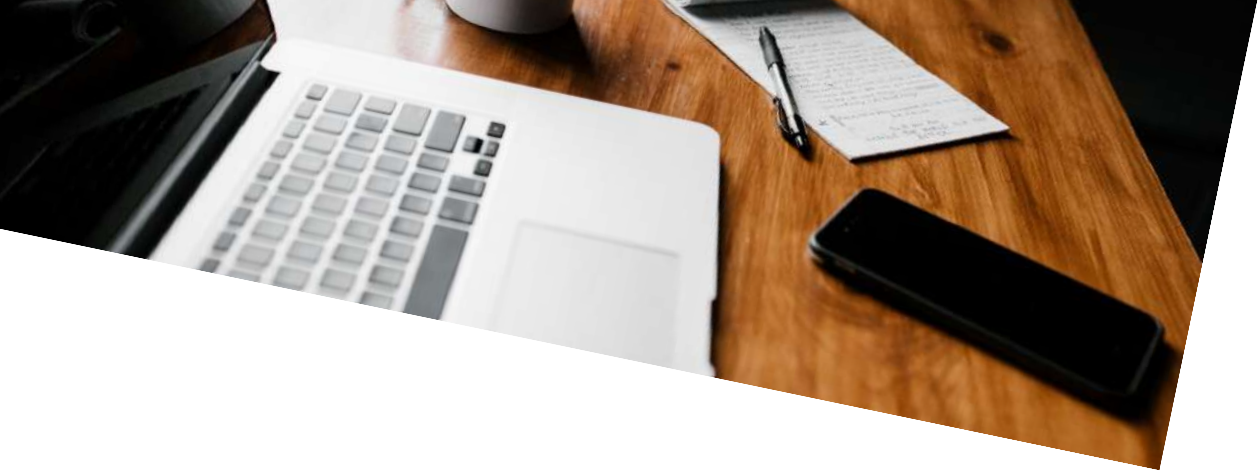

The Public Policy Programme at The Alan Turing Institute was set up in May 2018 with the aim of developing research, tools, and techniques that help governments innovate with data-intensive technologies and improve the quality of people's lives. We work alongside policy makers to explore how data science and artificial intelligence can inform public policy and improve the provision of public services. We believe that governments can reap the benefits of these technologies only if they make considerations of ethics and safety a first priority.

This workbook provides a summary of the end-to-end guidance on how to apply principles and best practices of AI explainability, *Explaining decisions made with AI,* a document co-produced by the Information Commissioner's Office and The Alan Turing Institute. The reader is referred to the full guidance for expansions of the concepts covered in this shorter intervention. The full guidance can be found here: https://ico.org.uk/for-organisations/guide-to-data-protection/key-data-protection-themes/explaining-decisions-made-with-ai/

Please note, that this workbook is a living document that will evolve and improve with input from users, affected stakeholders, and interested parties. We need your participation. Please share feedback with us at policy@turing.ac.uk

This work was supported exclusively by the Turing's Public Policy Programme under a research grant from UKRI's Strategic Priorities Fund (TU/ASG/R-SPEJ-102) as part of its AI for science and government programme.

https://www.turing.ac.uk/research/research-programmes/public-policy

Cite this work as:

Leslie, D., and Briggs, M. (2021). Explaining decisions made with AI: A workbook (Use case 1: AI-assisted recruitment tool). *The Alan Turing Institute*.

# Contents

## Explaining Decisions Made With AI: A Workbook

# INTRODUCTION

As artificial intelligence (AI) systems and machine learning (ML) techniques have advanced over last two decades, there has been a growing acknowledgement that these technologies hold an unprecedented promise to transform the world for the better. From revolutionising medical care, public health, environmental conservation, and energy consumption to accelerating scientific insights, improving agriculture, and helping governments address societal inequities, AI/ML applications are opening up a veritably unbounded range of potential benefits.

At the same time, however, hopes about the realisation of this windfall of possibilities have been accompanied by reasonable trepidations. Much of the progress that has been made in the development of ever more powerful AI/ML systems has depended upon a corresponding explosion of information processing power and an exponential expansion of the availability of large volumes of computable data. With more processing power and more data at their disposal, AI researchers and technologists have been able to design AI/ML models that are orders of magnitude more complex than the traditional rules-based systems that dominated the field of AI for its first half century. Many contemporary AI/ML applications, such as those in natural language processing and computer vision, complete their assigned tasks by identifying subtle patterns in vast datasets. These systems accomplish this by mathematically linking together many hundreds, thousands, millions—or sometimes even billions—of data points at a time. Humans don't think this way and because of this have difficulty understanding and explaining how these sorts of AI systems reach their results.

This gap in AI explainability becomes crucial when the outcomes of AI-assisted decisions have a significant impact on affected individuals and their communities. If an AI system is opaque then there is no way to ensure that its data processing is robust, reliable and safe. Similarly, in cases where social or demographic data are being used as inputs in AI decision-support systems—for instance, in domains such as criminal justice, social care, or job recruitment—the employment of 'black box' models leaves designers and deployers no way to properly safeguard against possibilities of lurking biases that may produce inequitable or discriminatory results.

Over the last two years, The Alan Turing Institute and the Information Commissioner's Office (ICO) have been working together to discover ways to tackle these difficult issues. The ultimate product of this joint endeavour, *Explaining decisions made with AI*, published in May 2020, is the most comprehensive practical guidance on AI explanation produced anywhere to date. We have put together this workbook to help support the uptake of that guidance.

The goal of the workbook is to summarise some of main themes from *Explaining decisions made with AI* and then to provide the materials for a workshop exercise that has been built around a use case created to help you gain a flavour of how to put the guidance into practice. In the first three sections, we run through the basics of *Explaining decisions made with AI*. We provide a precis of the four principles of AI explainability, the typology of AI explanations, and the tasks involved in the explanation-aware design, development, and use of AI/ML systems. We then provide some reflection questions, which are intended to be a launching pad for group discussion, and a starting point for the case-study-based exercise that we have included as Appendix B. In Appendix A, we go into more detailed suggestions about how to organise the workshop. These recommendations are based on two workshops we had the privilege of co-hosting with our colleagues from the ICO and Manchester Metropolitan University in January 2021. The participants of these workshops came from both the private and public sectors, and we are extremely grateful to them for their energy, enthusiasm, and tremendous insight. This workbook would simply not exist without the commitment and keenness of all our collaborators and workshop participants.

## How to Use this Workbook

This workbook contains summary information from *Explaining Decisions Made with AI*. In addition to giving a landscape view of the guidance, we have included a case study at the end of this document that we hope will bring the principles and recommendations it explores to life. We recommend following the tasks below in order to make full use of the materials in this workbook:
- Read the sections entitled "Four Principles of AI Explainability," "Explanation-Aware Design," and "Six Tasks for Examining AI in Practice"
- Consider the reflection questions and prepare some brief notes on your answers
- Read the Case Study presented in Appendix B which will form the basis for the workshop. Feel free to make any notes that you feel are relevant to the application of the guidance.
- Have awareness of the content of the Case Study Checklist found in Appendix C.
- Run through the Case Study in a group environment, following the procedure and timelines recommended in Appendix A.

This guidance is not a statutory code of practice under the Data Protection Act 2018 (DPA 2018) or otherwise. Instead, we aim to provide practically useful information about the ethics and governance of the explanation-aware design and use of AI that will help you engage in responsible research and innovation, build public trust, and demonstrate 'best practices'.

# FOUR PRINCIPLES OF AI EXPLAINABILITY

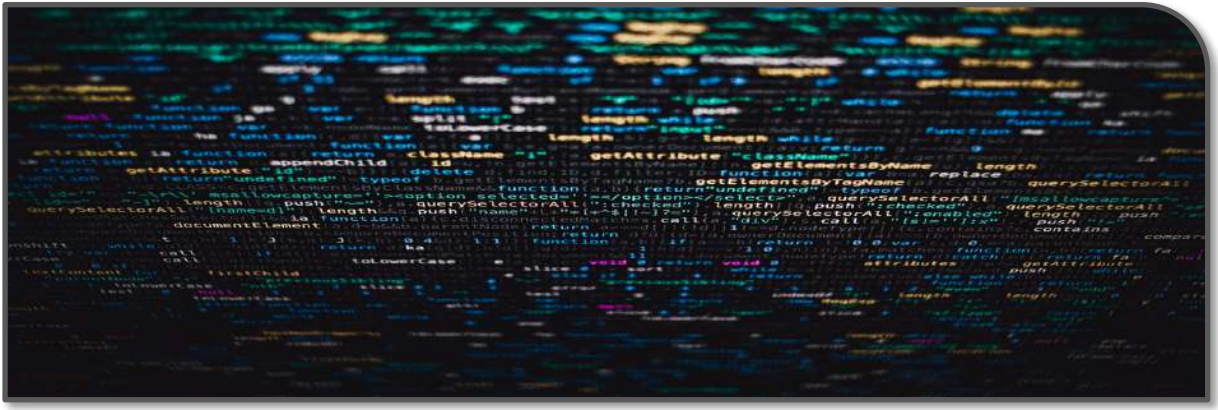

> " 
> *AI-assisted decisions are not unique to one sector, or to one type of organisation. They are increasingly used in all areas of life. This guidance recognises this, so you can use it no matter what your organisation does. The principles-based approach of this guidance gives you a broad steer on what to think about when explaining AI-assisted decisions to individuals.*

|  | Be transparent |
|---|---|
|  | Be accountable |
|  | Consider context |
|  | Reflect on impacts |

## Be transparent

*What is this principle about?*

The principle of being transparent is an extension of the transparency aspect of principle (a) in the GDPR (lawfulness, fairness and transparency).

In data protection terms, transparency means being open and honest about who you are, and how and why you use personal data.

Being transparent about AI-assisted decisions builds on these requirements. It is about making your use of AI for decision-making obvious and appropriately explaining the decisions you make to individuals in a meaningful way.

*What are the key aspects of being transparent?*

Raise awareness:

- Be open and candid about:
    - your use of AI-enabled decisions;
    - when you use them; and
    - why you choose to do this.
- Proactively make people aware of a specific AI-enabled decision concerning them, in advance of making the decision.

Meaningfully explain decisions:

- Don't just give **any** explanation to people about AI-enabled decisions - give them:
    - a truthful and meaningful explanation;
    - written or presented appropriately; and
    - delivered at the right time.

## Be accountable

*What is this principle about?*

The principle of being accountable is derived from the accountability principle in the GDPR.

In data protection terms, accountability means taking responsibility for complying with the other data protection principles and being able to demonstrate that compliance. It also means implementing appropriate technical and organisational measures, and data protection by design and default.

Being accountable for explaining AI-assisted decisions concentrates these dual requirements on the processes and actions you carry out when designing (or procuring/ outsourcing) and deploying AI models.

It is about ensuring appropriate oversight of your AI decision systems, and being answerable to others in your organisation, to external bodies such as regulators, and to the individuals you make AI-assisted decisions about.

*What are the key aspects of being accountable?*

Assign responsibility:

- Identify those within your organisation who manage and oversee the 'explainability' requirements of an AI decision system and assign ultimate responsibility for this.
- Ensure you have a designated and capable human point of contact for individuals to query or contest a decision.

Justify and evidence:

- Actively consider and make justified choices about how to design and deploy AI models that are appropriately explainable to individuals.
- Take steps to prove that you made these considerations, and that they are present in the design and deployment of the models themselves.
- Show that you provided explanations to individuals.

## Consider context

*What is this principle about?*

There is no one-size-fits-all approach to explaining AI-assisted decisions. The principle of considering context underlines this.

It is about paying attention to several different, but interrelated, elements that can have an effect on explaining AI-assisted decisions and managing the overall process.

This is not a one-off consideration. It's something you should think about at all stages of the process, from concept to deployment and presentation of the explanation to the decision recipient.

*What are the key aspects of considering context?*

*Choose appropriate models and explanation*

When planning on using AI to help make decisions about people, you should consider:

- the setting in which you will do this;
- the potential impact of the decisions you make;

- what an individual should know about a decision, so you can choose an appropriately explainable AI model; and
- prioritising delivery of the relevant explanation types.

*Tailor governance and explanation*

Your governance of the 'explainability' of AI models should be:

- robust and reflective of best practice; and
- tailored to your organisation and the particular circumstances and needs of each decision recipient.

# Reflect on impacts

## What is this principle about?

In making decisions and performing tasks that have previously required the thinking and reasoning of responsible humans, AI systems are increasingly serving as trustees of human decision-making. However, individuals cannot hold these systems directly accountable for the consequences of their outcomes and behaviours.

The value of reflecting on the impacts of your AI system helps you explain to individuals affected by its decisions that the use of AI will not harm or impair their wellbeing.

This means asking and answering questions about the ethical purposes and objectives of your AI project at the initial stages.

You should then revisit and reflect on the impacts identified in the initial stages of the AI project throughout the development and implementation stages. If any new impacts are identified, you should document them, alongside any mitigating factors you implement where relevant. This will help you explain to decision recipients what impacts you have identified and how you have reduced any potentially harmful effects as far as possible.

## What are the key aspects of reflecting on impacts?

*Individual wellbeing*

Think about how to build and implement your AI system in a way that:

- fosters the physical, emotional and mental integrity of affected individuals;
- ensures their abilities to make free and informed decisions about their own lives;
- safeguards their autonomy and their power to express themselves;

- supports their abilities to flourish, to fully develop themselves, and to pursue their interests according to their own freely determined life plans;
- preserves their ability to maintain a private life independent from the transformative effects of technology; and
- secures their capacities to make well-considered, positive and independent contributions to their social groups and to the shared life of the community, more generally.

*Wellbeing of wider society*

Think about how to build and implement your AI system in a way that:

- safeguards meaningful human connection and social cohesion;
- prioritises diversity, participation and inclusion;
- encourages all voices to be heard and all opinions to be weighed seriously and sincerely;
- treats all individuals equally and protects social equity;
- uses AI technologies as an essential support for the protection of fair and equal treatment under the law;
- utilises innovation to empower and to advance the interests and well-being of as many individuals as possible; and
- anticipates the wider impacts of the AI technologies you are developing by thinking about their ramifications for others around the globe, for the biosphere as a whole and for future generations.

*\*\*[See Part 1: "The Principles to Follow"](#) in the Guidance for further details\*\**

# EXPLANATION-AWARE DESIGN

There are different ways of explaining AI decisions. The table below illustrates the six main types of explanation and is followed by more granular descriptions of each of the explanation types.

| Six main types of explanation | Description |
|---|---|
| Rationale explanation | The reasons that led to a decision, delivered in an accessible and non-technical way. |
| Responsibility explanation | Who is involved in the development, management and implementation of an AI system, and who to contact for a human review of a decision. |
| Data explanation | What data has been used in a particular decision and how. |
| Fairness explanation | Steps taken across the design and implementation of an AI system to ensure that the decisions it supports are generally unbiased and fair, and whether or not an individual has been treated equitably. |
| Safety and performance explanation | Steps taken across the design and implementation of an AI system to maximise the accuracy, reliability, security and robustness of its decisions and behaviours. |
| Impact explanation | Steps taken across the design and implementation of an AI system to consider and monitor the impacts that the use of an AI system and its decisions has or may have on an individual, and on wider society. |

## Process-based vs outcome-based explanations

Before we explore the six explanation types, it is useful to make a distinction, which applies to all of them, between process-based and outcome-based explanations.

The primary aim of explaining fully automated or AI-assisted decisions is justifying a particular result to the individual whose interests are affected by it. This means:

- demonstrating how you and all others involved in the development of your system acted responsibly when choosing the processes behind its design and deployment; and
- making the reasoning behind the outcome of that decision clear.

We have therefore divided each type of explanation into the subcategories of 'process' and 'outcome':

- **Process-based explanations of AI systems** are about demonstrating that you have followed good governance processes and best practices throughout your design and use.

  For example, if you are trying to explain the fairness and safety of a particular AI-assisted decision, one component of your explanation will involve establishing that you have taken adequate measures across the system's production and deployment to ensure that its outcome is fair and safe.

- **Outcome-based explanations of AI systems** are about clarifying the results of a specific decision. They involve explaining the reasoning behind a particular algorithmically-generated outcome in plain, easily understandable, and everyday language.

  If there is meaningful human involvement in the decision-making process, you also have to make clear to the affected individual how and why a human judgement that is assisted by an AI output was reached.

  In addition, you may also need to confirm that the actual outcome of an AI decision meets the criteria that you established in your design process to ensure that the AI system is being used in a fair, safe, and ethical way.

## Rationale explanation

*What does this explanation help people to understand?*

It is about the 'why?' of an AI decision. It helps people understand the reasons that led to a decision outcome, in an accessible way.

*What you may need to show*

- How the system performed and behaved to get to that decision outcome.
- How the different components in the AI system led it to transform inputs into outputs in a particular way, so you can communicate which features, interactions, and parameters were most significant.
- How these technical components of the logic underlying the result can provide supporting evidence for the decision reached.
- How this underlying logic can be conveyed as easily understandable reasons to decision recipients.
- How you have thought about how the system's results apply to the concrete context and life situation of the affected individual.

*Rationale explanations might answer:*

- Have we selected an algorithmic model, or set of models, that will provide a degree of interpretability that corresponds with its impact on affected individuals?
- Are the supplementary explanation tools that we are using to help make our complex system explainable good enough to provide meaningful and accurate information about its underlying logic?

*Process-based explanations clarify:*

- How the procedures you have set up help you provide meaningful explanations of the underlying logic of your AI model's results.
- How these procedures are suitable given the model's particular domain context and its possible impacts on the affected decision recipients and wider society.
- How you have set up your system's design and deployment workflow so that it is appropriately interpretable and explainable, including its data collection and pre-processing, model selection, explanation extraction, and explanation delivery procedures.

*Outcome-based explanations provide:*

- The formal and logical rationale of the AI system – how the system is verified against its formal specifications, so you can verify that the AI system will operate reliably and behave in accordance with its intended functionality.
- The technical rationale of the system's output – how the model's components (its variables and rules) transform inputs into outputs, so you know what role these components play in producing that output. By understanding the roles and functions of the individual components, it is possible to identify the features and parameters that significantly influence a particular output.
- Translation of the system's workings – its input and output variables, parameters and so on – into accessible everyday language, so you can clarify, in plain and understandable terms, what role these factors play in reasoning about the real-world problem that the model is trying to address or solve.
- Clarification of how a statistical result is applied to the individual concerned. This should show how the reasoning behind the decision takes into account the specific circumstances, background and personal qualities of affected individuals.

# Responsibility explanation

*What does this explanation help people to understand?*

It helps people understand 'who' is involved in the development and management of the AI model, and 'who' to contact for a human review of a decision.

*What you may need to show*

- Who is accountable at each stage of the AI system's design and deployment, from defining outcomes for the system at its initial phase of design, through to providing the explanation to the affected individual at the end.
- Definitions of the mechanisms by which each of these people will be held accountable, as well as how you have made the design and implementation processes of your AI system traceable and auditable.

*Process-based explanations clarify:*

- The roles and functions across your organisation that are involved in the various stages of developing and implementing your AI system, including any human involvement in the decision-making. If your system, or parts of it, are procured, you should include information about the providers or developers involved.
- Broadly, what the roles do, why they are important, and where overall responsibility lies for management of the AI model – who is ultimately accountable.
- Who is responsible at each step from the design of an AI system through to its implementation to make sure that there is effective accountability throughout.

*Outcome-based explanations:*

Because a responsibility explanation largely has to do with the governance of the design and implementation of AI systems, it is, in a strict sense, entirely process-based. Even so, there is important information about post-decision procedures that you should be able to provide:

- Cover information on how to request a human review of an AI-enabled decision or object to the use of AI, including details on who to contact, and what the next steps will be (e.g., how long it will take, what the human reviewer will take into account, how they will present their own decision and explanation).
- Give individuals a way to directly contact the role or team responsible for the review. You do not need to identify a specific person in your organisation. One person involved in this should have implemented the decision and used the statistical results of a decision-support system to come to a determination about an individual.

## Data explanation

*What does this explanation help people to understand?*

Data explanations are about the 'what' of AI-assisted decisions. They help people understand what data about them, and what other sources of data, were used in a particular AI decision. Generally, they also help individuals understand more about the data used to train and test the AI model. You could provide some of this information within the fair processing notice you are required to provide under Articles 13 and 14 of the GDPR.

*What you may need to show*

- How the data used to train, test, and validate your AI model was managed and utilised from collection through processing and monitoring.
- What data you used in a particular decision and how.

*Process-based explanations include:*

- What training/ testing/ validating data was collected, the sources of that data, and the methods that were used to collect it.
- Who took part in choosing the data to be collected or procured and who was involved in its recording or acquisition. How procured or third-party provided data was vetted.
- How data quality was assessed and the steps that were taken to address any quality issues discovered, such as completing or removing data.
- What the training/ testing/ validating split was and how it was determined.
- How data pre-processing, labelling, and augmentation supported the interpretability and explainability of the model.
- What measures were taken to ensure the data used to train, test, and validate the system was representative, relevant, accurately measured, and generalisable.
- How you ensured that any potential bias and discrimination in the dataset have been mitigated.

*Outcome-based explanations:*

- Clarify the input data used for a specific decision, and the sources of that data. This is outcome-based because it refers to your AI system's result for a particular decision recipient.

- In some cases, the output data may also require an explanation, particularly where the decision recipient has been placed in a category which may not be clear to them. For example, in the case of anomaly detection for financial fraud identification, the output might be a distance measure which places them at a certain distance away from other people based on their transaction history. Such a classification may require an explanation.

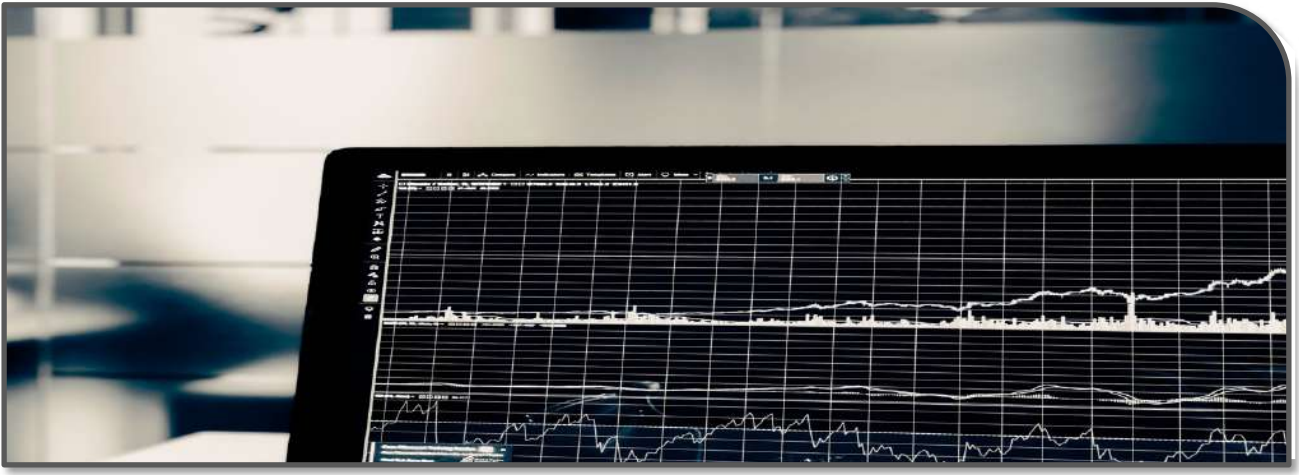

## Fairness explanation

*What does this explanation help people to understand?*

The fairness explanation is about helping people understand the steps you took (and continue to take) to ensure your AI decisions are generally unbiased and equitable. It also gives people an understanding of whether or not they have been treated equitably themselves.

*What you may need to show*

An explanation of fairness can relate to several stages of the design, development and deployment of AI systems:

**Dataset fairness:** The system is trained and tested on properly representative, relevant, accurately measured, and generalisable datasets (note that this dataset fairness component will overlap with data explanation). This may include showing that you have made sure your data is:

- as representative as possible of all those affected;
- sufficient in terms of its quantity and quality, so it represents the underlying population and the phenomenon you are modelling;

- assessed and recorded through suitable, reliable and impartial sources of measurement and has been sourced through sound collection methods;
- up-to-date and accurately reflects the characteristics of individuals, populations and the phenomena you are trying to model; and
- relevant by calling on domain experts to help you understand, assess and use the most appropriate sources and types of data to serve your objectives.

**Design fairness:** It has model architectures that do not include target variables, features, processes, or analytical structures (correlations, interactions, and inferences) which are unreasonable or unjustifiable. This may include showing that you have done the following:

- Attempted to identify any underlying structural biases that may play a role in translating your objectives into target variables and measurable proxies. When defining the problem at the start of the AI project, these biases could influence what system designers expect target variables to measure and what they statistically represent.
- Mitigated bias in the data pre-processing phase by taking into account the sector or organisational context in which you are operating. When this process is automated or outsourced, show that you have reviewed what has been done, and maintained oversight. You should also attach information on the context to your metadata, so that those coming to the pre-processed data later on have access to the relevant properties when they undertake bias mitigation.
- Mitigated bias when the feature space was determined (i.e., when relevant features were selected as input variables for your model). Choices made about grouping or separating and including or excluding features, as well as more general judgements about the comprehensiveness or coarseness of the total set of features, may have consequences for protected groups of people.
- Mitigated bias when tuning parameters and setting metrics at the modelling, testing and evaluation stages (i.e., into the trained model). Your AI development team should iterate the model and peer review it to help ensure that how they choose to adjust the dials and metrics of the model are in line with your objectives of mitigating bias.
- Mitigated bias by watching for hidden proxies for discriminatory features in your trained model, as these may act as influences on your model's output. Designers should also look into whether the significant correlations and inferences determined by the model's learning mechanisms are justifiable.

**Outcome fairness:** It does not have discriminatory or inequitable impacts on the lives of the people it affects. This may include showing that:

- you have been explicit about the formal definition(s) of fairness you have chosen and why. Data scientists can apply different formalised fairness criteria to choose how specific groups in a selected set will receive benefits in comparison to others in the same set, or how the accuracy or precision of the model will be distributed among subgroups; and
- the method you have applied in operationalising your formalised fairness criteria, for example, by reweighting model parameters; embedding trade-offs in a classification procedure; or re-tooling algorithmic results to adjust for outcome preferences.

**Implementation fairness:** It is deployed by users sufficiently trained to implement it responsibly and without bias. This may include showing that you have appropriately prepared and trained the implementers of your system to:

- avoid automation bias (over-relying on the outputs of AI systems) or automation-distrust bias (under-relying on AI system outputs because of a lack of trust in them);
- use its results with an active awareness of the specific context in which they are being applied. They should understand the particular circumstances of the individual to which that output is being applied; and
- understand the limitations of the system. This includes understanding the statistical uncertainty associated with the result as well as the relevant error rates and performance metrics.

*Process-based explanations include:*

- your chosen measures to mitigate risks of bias and discrimination at the data collection, preparation, model design and testing stages;
- how these measures were chosen and how you have managed informational barriers to bias-aware design such as limited access to data about protected or sensitive traits of concern; and
- the results of your initial (and ongoing) fairness testing, self-assessment, and external validation – showing that your chosen fairness measures are deliberately and effectively being integrated into model design. You could do this by showing that different groups of people receive similar outcomes, or that protected characteristics have not played a factor in the results.

*Outcome-based explanations include:*

- details about how your formal fairness criteria were implemented in the case of a particular decision or output;

- presentation of the relevant fairness metrics and performance measurements in the delivery interface of your model. This should be geared to a non-technical audience and done in an easily understandable way; and
- explanations of how others similar to the individual were treated (i.e., whether they received the same decision outcome as the individual). For example, you could use information generated from counter-factual scenarios to show whether or not someone with similar characteristics, but of a different ethnicity or gender, would receive the same decision outcome as the individual.

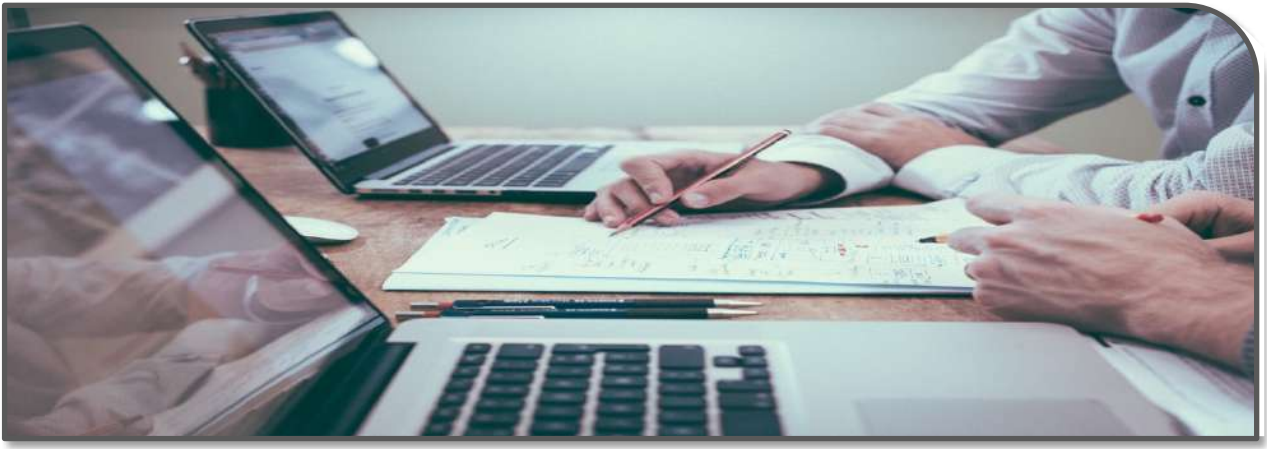

## Safety and performance explanation

*What does this explanation help people to understand?*

The safety and performance explanation helps people understand the measures you have put in place, and the steps you have taken (and continue to take) to maximise the accuracy, reliability, security and robustness of the decisions your AI model helps you to make. It can also be used to justify the type of AI system you have chosen to use, such as comparisons to other systems or human decision makers.

*What you may need to show*

**Accuracy:** the proportion of examples for which your model generates a correct output. This component may also include other related performance measures such as precision, sensitivity (true positives), and specificity (true negatives). Individuals may want to understand how accurate, precise, and sensitive the output was in their particular case.

**Reliability:** how dependably the AI system does what it was intended to do. If it did not do what it was programmed to carry out, individuals may want to know why, and whether this happened in the process of producing the decision that affected them.

**Security:** the system is able to protect its architecture from unauthorised modification or damage of any of its component parts; the system remains continuously functional and accessible to its authorised users and keeps confidential and private information secure, even under hostile or adversarial conditions.

**Robustness:** the system functions reliably and accurately in practice. Individuals may want to know how well the system works if things go wrong, how this has been anticipated and tested, and how the system has been immunised from adversarial attacks.

*Process-based explanations include:*

For accuracy:

- How you measure it (e.g., maximising precision to reduce the risk of false negatives).
- Why you chose those measures, and how you went about assuring it.
- What you did at the data collection stage to ensure your training data was up-to-date and reflective of the characteristics of the people to whom the results apply.
- What kinds of external validation you have undertaken to test and confirm your model's 'ground truth'.
- What the overall accuracy rate of the system was at testing stage.
- What you do to monitor this (e.g., measuring for concept drift over time).

For reliability:

- How you measure it and how you went about assuring it.
- Results of the formal verification of the system's programming specifications, i.e., how encoded requirements have been mathematically verified.

For security:

- How you measure it and how you went about assuring it, e.g., how limitation have been set on who is able to access the system, when, and how.
- How you manage the security of confidential and private information that is processed in the model.

For robustness:

- How you measure it.
- Why you chose those measures.
- How you went about assuring it, e.g., how you've stress-tested the system to understand how it responds to adversarial intervention, implementer error, or skewed goal-execution by an automated learner (in reinforcement learning applications).

*Outcome-based explanations:*

While you may not be able to guarantee accuracy at an individual level, you should be able to provide assurance that, at run-time, your AI system operated reliably, securely, and robustly for a specific decision.

- In the case of accuracy and the other performance metrics, however, you should include in your model's delivery interface the results of your cross-validation (training/ testing splits) and any external validation carried out.
- You may also include relevant information related to your system's confusion matrix (the table that provides the range of performance metrics) and ROC curve (receiver operating characteristics)/ AUC (area under the curve). Include guidance for users and affected individuals that makes the meaning of these measurement methods, and specifically the ones you have chosen to use, easily accessible and understandable. This should also include a clear representation of the uncertainty of the results (e.g., confidence intervals and error bars).

## Impact explanation

*What does this explanation help people to understand?*

An impact explanation helps people understand how you have considered the effects that your AI decision-support system may have on an individual, i.e., what the outcome of the decision means for them. It is also about helping individuals to understand the broader societal effects that the use of your system may have. This may help reassure people that the use of AI will be of benefit. Impact explanations are therefore often well suited to delivery before an AI-assisted decision has been made. See Task 6 of Explaining AI in practice for guidance on when to deliver explanations.

*What you may need to show*

- Demonstrate that you have thought about how your AI system will potentially affect individuals and wider society. Clearly show affected individuals the process you have gone through to determine these possible impacts.

*Process-based explanations include:*

- Showing the considerations you gave to your AI system's potential effects, how you undertook these considerations, and the measures and steps you took to mitigate possible negative impacts on society, and to amplify the positive effects.

- Information about how you plan to monitor and re-assess impacts while your system is deployed.

*Outcome-based explanations:*

Although the impact explanation is mainly about demonstrating that you have put appropriate forethought into the potential 'big picture' effects, you should also consider how to help decision recipients understand the impact of the AI-assisted decisions that specifically affect them. For instance, you might explain the consequences for the individual of the different possible decision outcomes and how, in some cases, changes in their behaviour would have brought about a different outcome with more positive impacts. This use of counterfactual assessment would help decision recipients make changes that could lead to a different outcome in the future or allow them to challenge the decision.

# SIX TASKS FOR THE EXPLANATION-AWARE DESIGN AND USE OF AI SYSTEMS

## Task 1: Select priority explanations by considering the domain, use case and impact on the individual

- Getting to know the different types of explanation will help you identify the dimensions of an explanation that decision recipients will find useful.
- In most cases, explaining AI-assisted decisions involves identifying what is happening in your AI system and who is responsible. That means you should prioritise the rationale and responsibility explanation types.
- The setting and sector you are working in is important in figuring out what kinds of explanation you should be able to provide. You should therefore consider domain context and use case.
- In addition, consider the potential impacts of your use of AI to determine which other types of explanation you should provide. This will also help you think about how much information is required, and how comprehensive it should be.
- Choosing what to prioritise is not an exact science, and while your choices may reflect what the majority of the people you make decisions about want to know, it's likely that other individuals will still benefit from the explanations you have not prioritised. These will probably also be useful for your own accountability or auditing purposes.

## Task 2: Collect and pre-process your data in an explanation-aware manner

- The data that you collect and pre-process before inputting it into your system has an important role to play in the ability to derive each explanation type.

- Careful labelling and selection of input data can help provide information for your rationale explanation.
- To be more transparent you may wish to provide details about who is responsible at each stage of data collection and pre-processing. You could provide this as part of your responsibility explanation.
- To aid your data explanation, you could include details on:
    - the source of the training data;
    - how it was collected;
    - assessments about its quality; and
    - steps taken to address quality issues, such as completing or removing data
- You should check the data used within your model to ensure it is sufficiently representative of those you are making decisions about. You should also consider whether pre-processing techniques, such as re-weighting, are required. These will help your fairness explanation.
- You should ensure that the modelling, testing and monitoring stages of your system development lead to accurate results to aid your safety and performance explanation.
- Documenting your impact and risk assessment, and steps taken throughout the model development to implement these assessments, will aid in your impact explanation.

## Task 3: Build your system to ensure you are able to extract relevant information for a range of explanation types

- Deriving the rationale explanation is key to understanding your AI system and helps you comply with parts of the GDPR. It requires looking 'under the hood' and helps you gather information you need for some of the other explanations, such as safety and performance and fairness. However, this is a complex task that requires you to know when to use more and less interpretable models and how to understand their outputs.
- To choose the right AI model for your explanation needs, you should think about the domain you are working in, and the potential impact of the deployment of your system on individuals and society.
- Following this, you should consider whether:
    - There are costs and benefits of replacing your current system with a newer and potentially less explainable AI model;
    - the data you use requires a more or less explainable system;
    - your use case and domain context encourage choosing an inherently interpretable system;
    - your processing needs lead you to select a 'black box' model; and
    - the supplementary interpretability tools that help you to explain a 'black box' model (if chosen) are appropriate in your context.
- To extract explanations from inherently interpretable models, look at the logic of the model's mapping function by exploring it and its results directly.

- To extract explanations from 'black box' systems, there are many techniques you can use. Make sure that they provide a reliable and accurate representation of the system's behaviour.

## Task 4: Translate the rationale of your system's results into useable and easily understandable reasons

- Once you have extracted the rationale of the underlying logic of your AI model, you will need to take the statistical output and incorporate it into your wider decision-making process.
- Implementers of the outputs from your AI system will need to recognise the factors that they see as legitimate determinants of the outcome they are considering.
- For the most part, the AI systems we consider in this guidance will produce statistical outputs that are based on correlation rather than causation. You therefore need to check whether the correlations that the AI model produces make sense in the case you are considering.
- Decision recipients should be able to easily understand how the statistical result has been applied to their particular case.

## Task 5: Prepare implementers to deploy your AI system

- In cases where decisions are not fully automated, implementers need to be meaningfully involved.
- This means that they need to be appropriately trained to use the model's results responsibly and fairly.
- Their training should cover:
    - the basics of how machine learning works;
    - the limitations of AI and automated decision-support technologies;
    - the benefits and risks of deploying these systems to assist decision-making, particularly how they help humans come to judgements rather than replacing that judgement; and
    - how to manage cognitive biases, including both decision-automation bias and automation-distrust bias.

## Task 6: Consider how to build and present your explanation

- To build an explanation, you should start by gathering together the information gained when implementing Tasks 1-4. You should review the information and determine how this provides an evidence base for the process-based or outcome-based explanations.
- You should then revisit the contextual factors to establish which explanation types should be prioritised.

- How you present your explanation depends on the way you make AI-assisted decisions, and on how people might expect you to deliver explanations you make without using AI.
- You can 'layer' your explanation by proactively providing individuals first with the explanations you have prioritised and making additional explanations available in further layers. This helps to avoid information (or explanation) overload.
- You should think of delivering your explanation as a conversation, rather than a one-way process. People should be able to discuss a decision with a competent human being.
- Providing your explanation at the right time is also important.
- To increase trust and awareness of your use of AI, you can proactively engage with your customers by making information available about how you use AI systems to help you make decisions.

**See Part 2 of the Guidance* for further details about each of the six tasks**

# REFLECTION QUESTIONS

- *What are your primary takeaways from the explanation-aware design and AI explainability principles?*
- *How feasible would it be for you to implement this guidance within your organisation?*
- *What are some technical or organisational barriers to applying the approach set out in the guidance within your organisation? How might these be overcome?*
- *How much importance do you currently place on the ethics knowledge and skills of yourself/your employees (in general, and specialists directly involved in governance processes)?*
- *Do you think that you/ your colleagues within your organisation have the necessary skills / training to be able to apply the guidance to your current / new AI related projects? If not, what would be the best model for delivery?*

# GLOSSARY

- **AI:** an umbrella term for a range of technologies and approaches that often attempt to mimic human thought to solve complex tasks.

- **Controller:** the individual or organisation which, alone or jointly with others, determines how and why personal data will be processed.

- **Data subject:** an identified or identifiable natural person, who can be identified, directly or indirectly, by information such as a name or identity number, or by a combination of characteristics specific to that individual.

- **Explainability:** level to which the resultant solutions of the applied technology can be understood by humans through clarification of the inner working of the model, leading to better decision-making and trust.

- **Generalisability:** an algorithm's ability to be effective across a wide range of inputs that reflect real world data.

- **'Human in the loop':** the outputs of an AI model can be used as part of a wider process in which a human considers the output, as well as other information available to them, and then acts (makes a decision) based on this.

- **Interpretability:** level to which a given algorithm makes sense to a user.

- **Legal or similarly significant effect:** something affecting an individual's legal status/ rights, or that has equivalent impact on an individual's circumstances, behaviour or opportunities, e.g., a decision about welfare, or a loan.

- **Linearity:** any change in the value of the predictor variable is directly reflected in a change in the value of the response variable at a constant rate. The interpretable prediction yielded by the model can therefore be directly inferred from the relative significance of the parameter/ weights of the predictor variable and have high inferential clarity and strength.

- **Monotonicity:** when the value of the predictor changes in a given direction, the value of the response variable changes consistently either in the same or opposite direction. The interpretable prediction yielded by the model can therefore be directly inferred. This monotonicity dimension is a highly desirable interpretability condition of predictive models in many heavily regulated sectors, because it incorporates reasonable expectations about the consistent application of sector specific selection constraints into automated decision-making systems.

- **Performance:** metrics used to define how good the algorithm is at reaching the correct decision that range beyond accuracy measures.

- **Personal data:** any information relating to a data subject.

- **Processing:** any operation or set of operations performed on personal data. These operations include: collecting, recording, organising, structuring, storing, adapting, retrieving or sharing of personal data.

- **Processor:** an individual or organisation which processes personal data on behalf of the controller.

- **Profiling:** any form of automated processing of personal data consisting of the use of personal data to evaluate certain characteristics of the data subject, in particular to analyse or predict aspects concerning that individual's work performance, health, economic situation, personal preferences, interests, behaviour or movements.

- **Solely automated processing:** where an automated system's output, and any decision taken as a result, are implemented without any human involvement or oversight.

- **Sparsity/ non-complexity:** the number of features (dimensionality) and feature interactions is low enough and the model of the underlying distribution is simple enough to enable a clear understanding of the function of each part of the model in relation to its outcome.

- **Supervised learning:** models are trained on a dataset which contains labelled data. 'Learning' occurs in these models when numerous examples are used to train an algorithm to map input variables (often called features) onto desired outputs (also called target variables or labels).

- **Third party:** an individual or organisation other than the data subject, controller or processor, or any individual or organisation authorised to process personal data on behalf of the data subject, controller or processor.

- **Training/testing/validation split:** process by which data set is divided into three subsets to estimate the model performance; the model is assessed on new data (data not used to train the model).

- **Unsupervised learning:** models are trained on a dataset without explicit instructions or labelled data. These models identify patterns and structures by measuring the densities or similarities of data points in the dataset.

# APPENDIX A: HOW TO USE THIS MATERIAL IN A WORKSHOP

There are two primary stages involved in using this workbook in a workshop setting, including pre-workshop activities and during the event. This layout is meant for online engagements, but it can easily be transitioned to in-person settings. A group of 15-25 participants is ideal for this type of workshop, thereby allowing for three breakout rooms of 5 to 9 participants; however, this is not necessary.

**Personnel needed:**

- One primary facilitator
- Three breakout room facilitators (if utilising breakout rooms)
- Three designated notetakers for each breakout room and discussion periods (one to two notetakers is still recommended regardless of breakout rooms)

**Pre-workshop:**

- Contact participants and send informed consent forms.
- Send out workbook material at least two to three days in advance of the workshop.
- Set up an e-board that can be used during the workshop.
- Prepare a slide deck for the initial presentation overview of the guidance and use case.
- Ensure all consent forms have been signed prior to the beginning of the workshop.

**During workshop:**

- Provide an overview of the Four Principles of AI Accountability and Explanation Types found in this guidance (e.g., 15-20 minute presentation).
- Provide some form of e-board medium where participants can post comments, responses to prompts, and any questions anonymously (e.g., [VWall](#)).
- Allow participants to engage on this medium for 15-20 minutes, responding to prompts such as:
    - *What are the main concerns raised by the AI-assisted HR recruitment tool for explanation-aware design?*
    - *Which explanation types would you choose in this case and how would you prioritise them?*
    - *What are some primary considerations for determining the model type in this case given the use-context and the potential impacts on affected individuals?*
- After this time has passed, the facilitator should draw out main themes and ask if any participants would like to elaborate on their comments.

- Depending on the group size, we would recommend splitting into breakout rooms for the next phase of the workshop. (E.g., a group of 20-25 can be divided into three groups of seven or eight). Each breakout room will focus on one of the explanation types laid out in the workbook. We have included a checklist for each explanation type as Appendix C. As means to practical exploration and discussion, each group should work through each of the points on the relevant checklist, running through scenarios of how the explainability guidance can be applied to the case study.
    - If there are not enough participants in the workshop to divide into breakout rooms, it is advisable to remain in one room together and focus instead on one or two of the explanation types.
- After the discussions in the breakout room, we recommend a 10-15 minute break for all participants.
- The next portion (50 minutes – 1 hour) of the workshop will be focused on reporting back discussions from the breakout rooms and/or discussing the applications of the chosen explanation types to the use case. This can be done by either asking a participant if they would like to provide a summary or having the facilitator report back main themes and ideas.
- The final 10 minutes are reserved for final questions.
- A post-survey should be sent around during the workshop, so that feedback can be gathered before participants leave.

**Possible workshop timelines:**

*20-25 participants*

| | |
|---|---|
| 9.30 – 9.50 | Presentation – Introduction to the Four Principles for AI Explainability and the Explanation Types |
| 9.50 – 10.05 | Introduction to Case Study |
| 10.05 – 10.35 | **PART 1:** Individual reflection and general group discussion using questions found on e-board<br>*15 minutes for reflection and 15 for discussion* |
| 10.35 – 11.20 | **PART 2:** Break out rooms<br>Group 1: Responsibility Explanation<br>Group 2: Data Explanation<br>Group 3: Fairness Explanation |
| 11.20 – 11.30 | Break |
| 11.30 – 12.20 | Discussion of use cases and feedback |
| 12.20 – 12.30 | Concluding remarks and post-workshop survey |

*10-20 participants*

| | |
|---|---|
| 1.30 – 1.50 | Presentation – Introduction to the Four Principles for AI Explainability and the Explanation Types |
| 1.50 – 2:15 | Introduction to Case Study |
| 2.15 – 3.00 | **PART 1:** Individual reflection and general group discussion using questions found on e-board<br>*15 minutes for reflection and 15 for discussion* |
| 3.00 – 3.15 | **Break** |
| 3.15 – 4.20 | PART 2: Group Discussion of 1-2 explanation types |
| 4.20 – 4.30 | Concluding remarks and post-workshop survey |

# APPENDIX B: CASE STUDY

## AI-assisted HR Recruitment Tool

Your company is considering using an AI-assisted recruitment tool to assist with future job vacancies by identifying certain traits and criteria to predict which candidates would be a good fit for your organisation. This could provide value and save time for your organisation as many talent acquisition leaders claim that the most difficult part of recruitment is identifying the right candidates from large applicant pools. It is becoming more and more difficult to read through the large volume of applications, and there is significant interest from your HR team to use an AI system to assist with this process. Your company generally receives hundreds of applications for each job posting, and an AI system presents itself as a tool to filter through applications and determine whether or not an applicant should receive an invitation to interview. The data available will consist of demographic data, general questions about skills, experience, and salary preferences, a CV, and an expression of interest; therefore, there are multiple data types that must be considered - both numeric and textual.

**Table 1**

*Table of Possible Features*

| Data Source | Feature | Data Type |
|---|---|---|
| Application question | Salary preference | Numeric |
| CV | Educational qualification(s) | Textual |
| CV | Years of work experience | Numeric |
| CV | Number of past roles | Numeric |
| CV | Previous job titles | Textual |
| Expression of Interest | Keyword search for terms of interest | Textual |
| Application question | List of skills (up to 5) | Textual |
| Application question | Have you been referred by someone within the organisation to apply to this position? | Textual |
| CV | Educational institution(s) attended | Textual |
| CV | Degree subject(s) | Textual |
| CV | Degree outcome (GPA, qualification, etc.) | Numeric/textual |
| Application question | Gender | Numeric |
| Application question | Race | Numeric |
| Application question | Ethnicity | Numeric |
| Application question | Prompt questions – keyword searches for terms of interest | Textual |

Demographic data will consist of numerically coded answers for questions of race, ethnicity, and gender. General questions about skills and experience will allow the applicant to fill in past work experience detailed in their CV, along with typed responses of up to 300 words in response to prompts such as, "Please describe a time in which you collaborated with stakeholders to design an effective research plan." The question regarding salary preference is also a free response question in which the applicant can type any numeric value. This cell will only recognise numbers not text; thus, the applicant must provide some numeric estimate in order to proceed with the application. The CV can be uploaded as a word document or a pdf file along with the expression of interest. Both documents will consist primarily of text. Applicants will upload and enter all of this information into our hiring platform, HIRED. The HR team can export this data to a .csv file with ease.

In addition to considering the possibility of using such a system, your team must also consider what type of model to choose. There are many available options including but not limited to regression, classification, and deep learning techniques such as neural networks. The graphic on the following page provides examples and further descriptions of each of these techniques. Linear regression is one of the most common modes of predictive analytics employed to model the relationship between one (or more) predictor variable(s) and one outcome variable. It is often seen as a desirable option, especially in high impact applications, because of its high degree of interpretability—its characteristics of linearity and monotonicity allow for output to be explainable and to meet reasonable expectations. On the other hand, a deep learning technique such as a supervised neural network, when given labelled data, uses multiple layers of nodes (weighted statistical variables) and a correction technique called backpropagation to adjust model weights in a way that minimises a cost function, so that the algorithm's predicted value is as close to the actual value as possible. This kind of algorithm is considered to be complex and opaque. It is a non-linear function that can have many variable interactions which are not human comprehensible.

Each of these algorithms have trade-offs in terms of performance, interpretability, and explainability. Your team must decide how to best balance the interpretability-accuracy trade-off. For example, a linear regression model is at a base level more interpretable than a neural network; however, when a linear regression model has more than two features, say 20 or 30, it automatically becomes increasingly complicated and less interpretable. One must recall also that explainability does not equate to fairness and that biases that may arise in any given model regardless of how explainable it has been designed to be. Your team must determine which model type is most appropriate for this specific use case.

| Linear Regression | Classification | Neural Network |
|---|---|---|
| 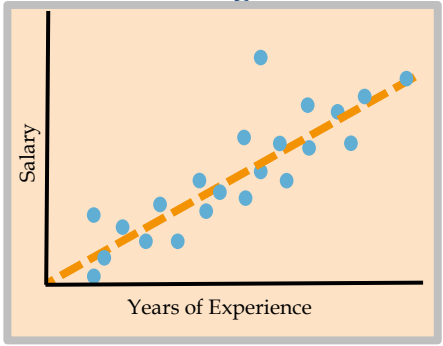 | 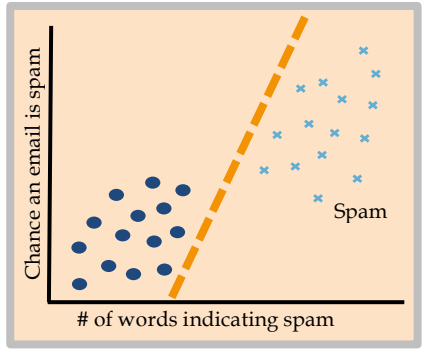 | 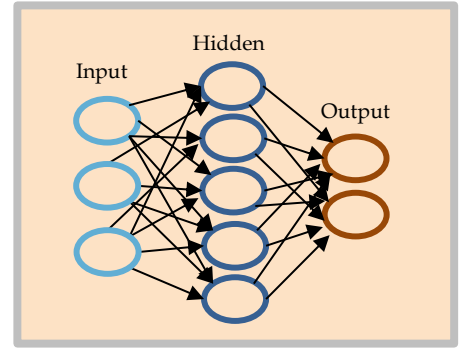 |
| Linear regression is a supervised learning technique in which input variables (often called features) are mapped onto desired outputs (also called target variables or labels). Linear regression is used to determine which input variables are best at predicting the value of a numeric target variable and estimate the magnitude and directional impact that the input variables have on the target variable. For example, linear regression could be used to determine if features such as years of experience and degree level have a significant relationship with salary across a specific sector. | Classification is another form of supervised learning in which input variables are mapped onto desired outputs. However, the primary difference between classification and regression is that the target variable is categorical in classification. The output can be binary such as yes or no, or multiclass which includes multiple categories. One of the most commonly used examples of classification is an email spam filter. A classification model uses various variables such as the presence of words like "lottery" or "you won" to predict whether or not an email should be classified as spam or not spam. | A supervised neural network is a technique that involves labelled target variables. By using multiple layers of nodes, the model minimises the cost function to determine what series of input weights will achieve the output with the least amount of error, so that the predicted value is as close to the actual value as possible. For example, an image classifier used to classify images into those with traffic lights and those without is given pre-labelled images (labelled as traffic lights or no traffic lights). The supervised neural network then determines the optimal input weights that minimise the overall error rate to classify the image as having traffic lights present or not. |

Additionally, your team may consider how textual data should be utilised. Will your team employ keyword searches, thematic coding, sentence semantic analysis, or a different form of Natural Language Processing such as sentiment analysis? Regardless of the model you choose, you plan to send an email to each applicant notifying them of their rejection or invitation to interview. This decision will be made by the AI system and fed into an automated response that can be personalised. In order to effectively train the model, you plan to use past applications to previous roles at your company. Your company has stored all previous applications from the last five years, creating a dataset of over 6,000 applications. This dataset is currently stored as a .csv file. One of your colleagues explains to you how the application process has changed over time to reflect industry standards; therefore, the data may not be standardised across the 6,000 applications.

Your team must now examine the potential use of an AI-assisted HR recruitment tool by applying the **explanation-aware design principles** along with other considerations from the *Explaining Decisions Made with AI* guidance. Here is a list of the project team members that may be involved at any point during the design, development, and use of the tool:

| Team member | Potential role and responsibilities |
| --- | --- |
| Product manager | The product manager defines product requirements for AI system, assure quality, and determines how it should be managed. They will be responsible throughout the AI lifecycle for ensuring the AI system is built to standard and maintained, and they are in charge seeing that improvements are made where needed. |
| Senior management | Senior management has the overall responsibility for ensuring the AI system that is developed and used within your organisation, or procured from a third party, is appropriately explainable to the decision recipient. Senior management and DPO should expect assurances from product manager that the system provides the appropriate level of explanation to applicants. |
| Domain expert in HR | Provides input to the AI development team including data wranglers and data scientists about how recruitment selection decisions are usually made. What qualities have been important to the organisation during recruitment processes in the past? Details how decisions were made and explained to applicants prior to using an AI system. |
| Data wrangler | The data wrangler collects, procures, and pre-processes the data that you input into your AI system, which must be representative, reliable, relevant, and up-to-date. |
| Data scientist | The data scientist builds and maintains the data analytic architecture and infrastructure that ensure the system performs as intended and that safeguard the extraction of appropriate technical explanations of the model. They also integrate the domain expertise shared by HR into explanation-aware design to ensure that the AI system is capable of delivering the types of explanations required. The data scientist builds, trains, and optimises the models; tests the model, deploys it, extracts explanations from it, and supports implementers in deploying it in practice. |
| Compliance officer/DPO | Ensures that the development and use of the AI system comply with regulations and an organisation's policies and governance procedures. This includes compliance with data protection law, such as the expectation that AI-assisted decisions are explained to individuals affected by those decisions, in this case, applicants. |
| Implementer in HR | Where there is a human in the loop (i.e., the decision is not fully automated) the implementer in HR will rely on the model developed to help complete the first phase of recruitment. When the AI-assisted recruitment tool is used to extract an explanation, implementers will either directly draw on the internal characteristics of the model (if it is inherently interpretable and simple) or use supplementary tools and methods that enable explanation, if it is not. If the system is developed by a third party, the implementers must have sufficient training and support, so that they fully understand the model and can deploy the system responsibly. |
| Hiring manager | In this case, the hiring manager will receive the output of the AI system. As your organisation has decided to send an email to all applicants, regardless of rejection or movement onto the interview stage, the hiring manager can personalise the email to the applicant. Additionally, the hiring manager should ensure that an accurate, context sensitive explanation can be delivered to the applicants regarding the outcome of their application, and that the decisions made with the aid of the AI system are equitable across applicants of different backgrounds. |

# APPENDIX C: CASE STUDY CHECKLIST

Please refer to the information contained in this workbook for more detail. For this case study, we will focus primarily on responsibility explanation, data explanation, and fairness explanation.

## Responsibility explanation

*What you may need to show*

☐ Who is accountable at each stage of the AI system's design and deployment, from defining outcomes for the system at its initial phase of design, through to providing the explanation to the affected individual at the end.

☐ Definitions of the mechanisms by which each of these people will be held accountable, as well as how you have made the design and implementation processes of your AI system traceable and auditable.

*What goes into the responsibility explanation*

☐ The roles and functions across your organisation that are involved in the various stages of developing and implementing your AI system, including any human involvement in the decision-making. If your system, or parts of it, are procured, you should include information about the providers or developers involved.

☐ Broadly, what the roles do, why they are important, and where overall responsibility lies for management of the AI model – who is ultimately accountable.

☐ Who is responsible at each step from the design of an AI system through to its implementation to make sure that there is effective accountability throughout.

☐ How to request a human review of an AI-enabled decision or object to the use of AI, including details on who to contact, and what the next steps will be (e.g., how long it will take, what the human reviewer will take into account, how they will present their own decision and explanation).

☐ How to directly contact the role or team responsible for the review. You do not need to identify a specific person in your organisation. One person involved in this should have implemented the decision and used the statistical results of a decision-support system to come to a determination about an individual.

# Data explanation

*What you may need to show*

☐ How the data used to train, test, and validate your AI model was managed and utilised from collection through processing and monitoring.

☐ What data you used in a particular decision and how.

*What goes into the data explanation*

☐ What training/ testing/ validating data was collected, the sources of that data, and the methods that were used to collect it.

☐ Who took part in choosing the data to be collected or procured and who was involved in its recording or acquisition. How procured or third-party provided data was vetted.

☐ How data quality was assessed and the steps that were taken to address any quality issues discovered, such as completing or removing data.

☐ What the training/ testing/ validating split was and how it was determined.

☐ How data pre-processing, labelling, and augmentation supported the interpretability and explainability of the model.

☐ What measures were taken to ensure the data used to train, test, and validate the system was representative, relevant, accurately measured, and generalisable.

☐ How you ensured that any potential bias and discrimination in the dataset have been mitigated.

☐ Clarification of the input data used for a specific decision, and the sources of that data.

☐ Clarification of the output data, particularly where the decision recipient has been placed in a category which may not be clear to them.

# Fairness explanation

*What you may need to show*

**Dataset fairness:** The system is trained and tested on properly representative, relevant, accurately measured, and generalisable datasets (note that this dataset fairness component will overlap with data explanation). This may include showing that you have made sure your data is:

☐ as representative as possible of all those affected;

☐ sufficient in terms of its quantity and quality, so it represents the underlying population and the phenomenon you are modelling;

☐ assessed and recorded through suitable, reliable and impartial sources of measurement and has been sourced through sound collection methods;

☐ up-to-date and accurately reflects the characteristics of individuals, populations and the phenomena you are trying to model; and

☐ relevant by calling on domain experts to help you understand, assess and use the most appropriate sources and types of data to serve your objectives.

**Design fairness:** It has model architectures that do not include target variables, features, processes, or analytical structures (correlations, interactions, and inferences) which are unreasonable or unjustifiable. This may include showing that you have done the following:

☐ Attempted to identify any underlying structural biases that may play a role in translating your objectives into target variables and measurable proxies. When defining the problem at the start of the AI project, these biases could influence what system designers expect target variables to measure and what they statistically represent.

☐ Mitigated bias in the data pre-processing phase by taking into account the sector or organisational context in which you are operating. When this process is automated or outsourced, show that you have reviewed what has been done, and maintained oversight. You should also attach information on the context to your metadata, so that those coming to the pre-processed data later on have access to the relevant properties when they undertake bias mitigation.

☐ Mitigated bias when the feature space was determined (i.e., when relevant features were selected as input variables for your model). Choices made about grouping or separating and including or excluding features, as well as more general judgements about the comprehensiveness or coarseness of the total set of features, may have consequences for protected groups of people.

☐ Mitigated bias when tuning parameters and setting metrics at the modelling, testing and evaluation stages (i.e., into the trained model). Your AI development team should iterate the model and peer review it to help ensure that how they choose to adjust the dials and metrics of the model are in line with your objectives of mitigating bias.

☐ Mitigated bias by watching for hidden proxies for discriminatory features in your trained model, as these may act as influences on your model's output. Designers should also look into whether the significant correlations and inferences determined by the model's learning mechanisms are justifiable.

**Outcome fairness:** It does not have discriminatory or inequitable impacts on the lives of the people it affects. This may include showing that:

☐ You have been explicit about the formal definition(s) of fairness you have chosen and why. Data scientists can apply different formalised fairness criteria to choose how specific groups in a selected set will receive benefits in comparison to others in the same set, or how the accuracy or precision of the model will be distributed among subgroups; and

☐ The method you have applied in operationalising your formalised fairness criteria, for example, by reweighting model parameters; embedding trade-offs in a classification procedure; or re-tooling algorithmic results to adjust for outcome preferences.

**Implementation fairness:** It is deployed by users sufficiently trained to implement it responsibly and without bias. This may include showing that you have appropriately prepared and trained the implementers of your system to:

☐ Avoid automation bias (over-relying on the outputs of AI systems) or automation-distrust bias (under-relying on AI system outputs because of a lack of trust in them);

☐ Use its results with an active awareness of the specific context in which they are being applied. They should understand the particular circumstances of the individual to which that output is being applied; and

☐ Understand the limitations of the system. This includes understanding the statistical uncertainty associated with the result as well as the relevant error rates and performance metrics.

*What goes into the fairness explanation*

☐ What measures were chosen to mitigate risks of bias and discrimination at the data collection, preparation, model design and testing stages;

☐ How these measures were chosen and how you have managed informational barriers to bias-aware design such as limited access to data about protected or sensitive traits of concern; and

☐ Results of your initial (and ongoing) fairness testing, self-assessment, and external validation – showing that your chosen fairness measures are deliberately and effectively being integrated into model design. You could do this by showing that different groups of people receive similar outcomes, or that protected characteristics have not played a factor in the results.

☐ Details about how your formal fairness criteria were implemented in the case of a particular decision or output;

☐ Presentation of the relevant fairness metrics and performance measurements in the delivery interface of your model. This should be geared to a non-technical audience and done in an easily understandable way; and

☐ Explanations of how others similar to the individual were treated (i.e., whether they received the same decision outcome as the individual).

The information contained in this workbook serves as preparation material that can assist with both participation and understanding for workshops involving the *Explaining Decisions Made with AI* Guidance.

The full link to this Guidance can be found at: https://ico.org.uk/for-organisations/guide-to-data-protection/key-data-protection-themes/explaining-decisions-made-with-ai/

This workbook design has been built on a template from usedtotech.com, Kawish, M. (2020). https://usedtotech.com/books/free-awesome-looking-workbook-design-in-microsoft-word/

Photo of people working (Title Page) – Taken by airfocus
Photo of code (Title Page) – Taken by Luis Gomes
Photo of phone, computer, and notebook (Page 2) – Taken by Andrew Neel
Photo of code (Page 6) – Taken by Markus Spiske
Photo of graph on screen (Page 16) – Taken by Chris Liverani
Photo of computer workspace (Page 19) – Taken by Scott Graham